# Lamb-Dicke spectroscopy of atoms in a hollow-core photonic crystal fibre

Shoichi Okaba[1,2], Tetsushi Takano[1,2], Fetah Benabid[3,4], Tom Bradley[3,4], Luca Vincetti[3,5], Zakhar Maizelis[6,7], Valery Yampol'skii[6,7], Franco Nori[8,9], & Hidetoshi Katori[1,2,10,11]

[1]Department of Applied Physics, Graduate School of Engineering, The University of Tokyo, 7-3-1 Bunkyo-ku, Tokyo 113-8656, Japan.

[2]Innovative Space-Time Project, ERATO, Japan Science and Technology Agency, 7-3-1 Bunkyo-ku, Tokyo 113-8656, Japan.

[3] GPPMM group, Xlim Research Institute, CNRS UMR7252, 123 av Albert Thomas, Limoges, France.

[4] Physics department, University of Bath, Claverton Down, Bath BA2 7AY, UK.

[5]Department of Engineering "Enzo Ferrari", University of Modena and Reggio Emilia, Modena I-41125, Italy.

[6]A.Ya. Usikov Institute for Radiophysics and Electronics, National Academy of Science of Ukraine, 12 Acad. Proskura Str., Kharkov 61085, Ukraine.

[7]V.N. Karazin Kharkov National University, 4 Pl. Svobody, Kharkov 61077, Ukraine.

[8] CEMS, RIKEN, 2-1 Hirosawa, Wako, Saitama 351-0198, Japan.

[9] Physics Department, University of Michigan, Ann Arbor, USA.

[10]Quantum Metrology Laboratory, RIKEN, 2-1 Hirosawa, Wako, Saitama 351-0198, Japan.

[11]*RIKEN Center for Advanced Photonics, 2-1 Hirosawa, Wako, Saitama 351-0198, Japan.*

**Abstract:**

**Unlike photons, which are conveniently handled by mirrors and optical fibres without loss of coherence, atoms lose their coherence via atom-atom and atom-wall interactions. This decoherence of atoms deteriorates the performance of atomic clocks and magnetometers, and also hinders their miniaturisation. Here we report a novel platform for precision spectroscopy. Ultracold strontium atoms inside a kagome-lattice hollow-core photonic crystal fibre are transversely confined by an optical lattice to prevent atoms from interacting with the fibre wall. By confining at most one atom in each lattice site, to avoid atom-atom interactions and Doppler effect, a 7.8-kHz-wide spectrum is observed for the $^1S_0 - {}^3P_1\,(m=0)$ transition. Atoms singly trapped in a magic lattice in hollow-core photonic crystal fibres improve the optical depth while preserving atomic coherence time.**

**Introduction**

Quantum metrology[1] with atoms relies on the long coherence times of atoms and photons to perform measurements at the quantum limit, which finds broad applications including atomic clocks[2], atom interferometers[3], magnetometers[4], and quantum simulators[5]. In particular, seeking minimally-perturbed optically-dense atomic samples[6] has been a long-standing endeavour in atomic, molecular and optical physics, as the quantum projection noise (QPN)[7] in the measurements reduces with the number of atoms as $N_a^{-1/2}$, which improves sensitivities in spectroscopy[8], optical magnetometry[4], and search for the permanent electric dipole moment (EDM)[9]. Since the atomic absorption cross section of a resonant light with wavelength $\lambda$ is given by $\sigma_0 = 3\lambda^2/2\pi$, small beam radius is preferable for efficient coupling[10] between atoms and photons. In a free space Gaussian beam, however, as the Rayleigh range changes as $z_R = \pi w_0^2/\lambda$, an effective interaction volume $V = \pi w_0^2 z_R$ rapidly decreases as the waist radius $w_0^4$. One then is confronted with severe trade-offs: a strong atom-photon interaction is available at the price of increased atom density $n = N_a/V$, which causes harmful atom-atom interactions. Atoms in a cavity have been used to get rid of this constraint by enabling multiple interaction[11,12] of photons $N_p$ with atoms $N_a$ inside a cavity mode volume $V \approx \pi w_0^2 L$ with $L$ the length of the cavity.

Atoms coupled to optical fibres may offer an alternative, as their effective interaction length can be arbitrarily longer than the Rayleigh range $z_R$. Strong coupling of spontaneous photons into the guided mode of a nanofibre was predicted and observed[13]. Evanescent trapping of atoms in the vicinity of the nanofibre allowed for an interface between atoms and guided modes[14]. A state-insensitive two-colour dipole trapping around the nanofibre was demonstrated[15] to reveal the natural linewidth of the $6S_{1/2}, F = 4 \to 6P_{3/2}, F' = 5$ transition of Cs by cancelling out the light shifts of trapping lasers. On the other hand, hollow-core photonic crystal fibres (HC-PCFs)[16] have been proven to confine thermal atoms or molecules together with the guided light inside its core over several meters, enabling the generation of optical nonlinearities at ultra-low optical power-levels, and sensitive spectroscopy on weak transitions[17]. For example, a $10$ kHz precision was demonstrated for a saturated absorption spectroscopy of acetylene by extrapolating the zero pressure-shift[18]. In order to prevent atoms from interacting with the fibre walls, optical dipole trapping of atoms in HC-PCFs was applied to guide thermal Rb gas[19], Bose condensed Na gas[20], and laser-cooled Rb gas[21]. Removal of the light shift perturbation during absorption spectroscopy was demonstrated by temporarily turning off a guiding laser[22].

However, in all the above configurations, the finest fibre-based spectroscopy that has so far been measured still exhibits a several-MHz-wide spectral linewidth. This broad linewidth is set by the strong collision of the confined thermal molecules or atoms[18,23] with the fibre core inner-wall, or by the natural linewidth of alkali atoms[15,22]. As such, the coherence of atoms or molecules longer than tens of ns in fibres remains a challenge. Targeting ultra-precision laser spectroscopy of atoms at a fractional uncertainty of $10^{-17}$ and beyond, which is regarded as a goal accuracy for next-generation atomic clocks[24,25], we considered possible fibre-based configurations. Depending on the atom-wall distance, from tens of nm to tens of µm, the atom fibre-wall interactions change from van der Waals, Casimir-Polder, and finally to thermal-bath regime[26]. The van der Waals interaction strongly shifts atomic lines by as much as $10^{-10}$ of fractional frequency-shift, for atoms trapped in the evanescent field[13-15] at tens of nm from the nanofibres. In order to keep the atom-wall-interaction-induced fractional frequency-shift below $10^{-17}$, one has to keep the atom-fibre walls distance $r_c \gtrsim 20$ µm (see Methods).

Here, we investigate the $^1S_0 - {}^3P_1$ transition of $^{88}$Sr atoms in a 1D optical lattice tuned to the magic condition, which confines atoms near the centre of the HC-PCF and in the Lamb-Dicke regime without introducing a light shift[15,24,27]. The moderately narrow linewidth $\gamma_p = 7.5$ kHz of the transition[28] offers an efficient probe to characterize atom-atom and atom-fibre interactions by its spectral line broadenings and shifts. Well-characterized collisional properties[29] allow investigating the occupancy of atoms in lattices through the reinforced collisional shift by the resonant dipole-dipole interaction, while the total angular momentum $J = 1$ of the upper state probes the fibre birefringence effects via the tensor light shift[27]. By carefully eliminating collisional shift and birefringence-induced light-shift, we show that the atomic resonance frequency can be unaffected by the fibre within an uncertainty of 0.11 kHz or $\approx 3 \times 10^{-13}$. Our investigations provide useful insight for designing fibre-based optical lattice clocks on the mHz-narrow $^1S_0 - {}^3P_0$ clock transitions[30], where both the collisional and polarization-dependent light shift are expected to be suppressed by more than 3-7 orders of magnitude, depending on the isotopes to be used[31].

**Results**

*Experimental setup:*

Figure 1a shows our experimental setup. $^{88}$Sr atoms are laser-cooled and trapped at a temperature of a few μK using a narrow line magneto-optical trap (MOT)[32]. A 32-mm-long kagome cladding lattice HC-PCF with hypocycloid core-shape[16,33,34] is placed near the MOT. The fibre covers the experimental wavelengths (689 – 914 nm) with loss figures of less than 650 dB km$^{-1}$ and guides dominantly in the HE$_{11}$ mode (see Figs. 1e-f and Methods). We couple optical lattice lasers at wavelength $\lambda_L \approx$ 813 nm from both ends of the HC-PCF. The potential depth of the optical lattice is about 30 μK at the MOT position ($z = -1.6$ mm from the entrance end of the HC-PCF) and 300 μK inside the fibre. After loading roughly $10^4$ atoms into the optical lattice, the atoms are adiabatically accelerated up to $v_\mathrm{m} \approx 53$ mm/s and transported inside the fibre hollow core to a position $z$. The atom acceleration and positioning is controlled by the frequency difference $\delta\nu(t) = \nu_2(t) - \nu_1$ of the lattice lasers, as outlined in Figs. 1a and 1b. For the adiabatic acceleration (deceleration) of atoms, we linearly chirp ($\approx 2$ kHz ms$^{-1}$) the frequency difference $\delta\nu(t)$ over 60 ms. The transport velocity $v_\mathrm{m}$ is optimised to maximize the number of atoms that pass through the 32-mm-long fibre by considering the trade-offs: while shorter transit time ($\sim v_\mathrm{m}^{-1}$) through the fibre reduces the collision losses, larger $v_\mathrm{m}$ increases the heating loss of atoms, as we discuss below.

*Lifetime of atoms in a fibre:*

The lifetime of the atoms in the fibre is of serious concern to discuss its potential applications, as glancing collisions with residual gases severely limit the coherence time of trapped atoms. The inset of Fig. 2 shows the lifetime $\tau = 347$ ms $\pm$ 8 ms of atoms trapped at $z = 23.4$ mm. Figure 2 maps out the position-dependent lifetime of atoms along the fibre. The lifetime of $\tau = 500$ ms near the entrance of the fibre, which is close to that measured outside the fibre, decreases to $\tau = 350$ ms in the middle of the fibre. Using a glancing-collision model[35] and taking into account the measured lifetimes and trap depth, we estimate the vacuum pressure in the middle and outside of the fibre to be $P_\mathrm{in} \approx 1.7 \times 10^{-6}$ Pa and $P_\mathrm{out} \approx 1 \times 10^{-6}$ Pa. The latter is in good agreement with the measured vacuum pressure, and the increase of the pressure in the fibre is reasonably accounted for by a small core radius $r_\mathrm{c} = 17$ μm of the fibre and by the outgassing rate $q$ per unit surface area of the fibre wall. By solving the 1D diffusion equation, we obtain the steady-state solution $(\partial P_\mathrm{PCF}/\partial t = 0)$ of the pressure inside the fibre as $P_\mathrm{PCF}(z, l) = -\frac{q}{r_\mathrm{c} D} z(z - l) + P_\mathrm{out}$, for $0 < z < l$, with $l$ being the length of the fibre and $D$ the diffusion constant. Based on the estimated pressure $P_{\text{PCF}}(l/2, l)$ with $l = 32$ mm, the pressure in the middle section of the fibre for an arbitrary length $l$ would scale as $P_{\text{PCF}}(l/2, l) \approx 7 \times 10^{-4} l^2$ Pa m$^{-2}$ + $P_{\text{out}}$. We expect that intensive baking of the fibre may reduce the outgassing rate $q$, thus extending the lifetimes of atoms for a longer fibre for future experiments.

The above result also suggests that there is no extra heating loss of atoms in the fibre as long as atoms are held at the same position. However, we observe larger heating of atoms, as the transport velocity $v_m$ increases. At $v_m = 53$ mm s$^{-1}$, the heating rate is estimated to be $\sim 300$ μK s$^{-1}$ for the moving lattice potential depth of 180 μK. We attribute this to a parametric heating of atoms caused by a residual standing-wave field, which is created by a partial reflection (~0.5 %) of the lattice laser by a viewport. This standing-wave potential modulates the moving lattice potential by 7 % at the frequency $f_m = 2v_m/\lambda_L \sim 130$ kHz, as atoms travel every $\lambda_L/2$. As $v_m$ increases, this frequency becomes closer to the parametric resonance condition[36] $f_m = 2f_{\text{lattice}}/n$, where $f_{\text{lattice}} \approx 300$ kHz is the vibrational frequency of the lattice and $n = 4$. To cope with this heating, we apply laser cooling during transport, which successfully reduces the heating loss of atoms.

*Absorption spectroscopy:*

We perform absorption spectroscopy for atoms trapped at $z \sim 4$ mm. The $^1S_0 - {}^3P_1(m = 0)$ intercombination transition at $\lambda_p = 689$ nm is probed by a laser, whose linewidth and frequency drift per hour are both less than 1 kHz by referencing a cavity made of ultra-low expansion (ULE) glass. We apply a bias field of $\mathbf{B}_0 = (0.14 \text{ mT})\hat{\mathbf{e}}_x$ perpendicular to the horizontal plane (see Fig. 1a) to define the quantization axis. The probe laser is linearly polarized with its electric field $\mathbf{E}_p$ parallel to $\mathbf{B}_0$, to excite the $\pi$ transition. The differential light shift for the transition is given by[27]

$$\Delta\nu_L = \Delta\tilde{\alpha}(\lambda_L, \boldsymbol{\epsilon}_L) I_L,$$

(1)

where $\Delta\tilde{\alpha}(\lambda_L, \boldsymbol{\epsilon}_L)$ is the differential polarizability, which depends on the lattice laser wavelength $\lambda_L$ and its polarization $\boldsymbol{\epsilon}_L$. The magic condition $\Delta\tilde{\alpha}(\lambda_L, \boldsymbol{\epsilon}_L) = 0$, to remove the differential light shift, can be satisfied for $690\text{ nm} < \lambda_L < 915\text{ nm}$ by tuning the tensor contribution of the light shift in the $^3P_1(m=0)$ state, which is determined by the angle $\theta_L$ of the linearly-polarized lattice laser $\boldsymbol{\epsilon}_L = \mathbf{E}_L/|\mathbf{E}_L|$ with respect to the quantization axis. It is noteworthy that, despite the fact that the HC-PCF guided mode exhibits a small longitudinal component $E_z$ (see methods), this is cancelled out in the standing-wave configuration.

We couple a probe intensity of $I_p \approx 0.15 I_0$ into the fibre, with $I_0 = 3\mu\text{W cm}^{-2}$, the saturation intensity of the transition. The transmission through atoms in the fibre is coupled to an avalanche photodiode (APD), as shown in Fig. 1a, where the overall photon-counting efficiency is estimated to be 30%. We define the frequency-dependent optical depth as

$$\text{OD}(\Delta\nu_p) = \frac{1}{1 + I_p/I_0 + (2\Delta\nu_p/\gamma_p)^2} \cdot \frac{2}{\pi w_0^2} \int_0^l dz \int_0^{r_c} dr\, 2\pi r n(z,r) \frac{3\lambda_p^2}{2\pi} e^{-2r^2/w_0^2}$$

where $\Delta\nu_p = \nu_p - \nu_0$ is the detuning of the probe laser. Here, we approximate the Bessel-mode profile of the guided mode to a Gaussian one with $w_0 = 11.8\ \mu\text{m}$ (see Methods), and $n(z,r) = \rho(z) e^{-r^2/w_a^2}$ assumes an atomic density distribution with $w_a \approx 2.0\ \mu\text{m}$, estimated from the atomic temperature and radial-trapping frequency of

≈ 1.3 kHz. Here, $l = 32$ mm and $r_c = 17$ μm are the length and hollow-core radius of the fibre, respectively. Using the photon counting rates with and without atoms, $\Pi_w$ and $\Pi_{w/o}$, and the background count rate $\Pi_{bk}$, the transmittance of the fibre is given by $T = \frac{\Pi_w - \Pi_{bk}}{\Pi_{w/o} - \Pi_{bk}}$, which is used to derive the optical depth as $OD(\Delta\nu_p) = -\ln T$. The number of atoms in the fibre is given by $N_a \approx 1200 \cdot OD(0)$. To avoid excess light shifts during spectroscopy, we reduce the lattice intensity by one order of magnitude from that used during the atom transfer. The probing time of the transition is limited to 3 ms, to reduce the photon-recoil heating loss of atoms out of the lattice potential.

*Collisional shift and its suppression:*

Figure 3a shows the measured optical depth $OD(\Delta\nu_p)$ as a function of the probe laser frequency. The Lamb-Dicke confinement and the light-shift cancellation allow us to approach the natural linewidth of the transition. However, as shown by the red symbols in Fig. 3b and 3c, we observe a collisional shift and broadening (see methods) for $OD(0) > 0.8$, which corresponds to the mean atom-occupation of each lattice site $\bar{m} = N_a \lambda_L/(2l_a) > 0.55$. Here, the atom cloud length $l_a$ is measured by the laser-induced fluorescence image of atoms, after extraction from the fibre by the moving lattice.

In order to make the high optical depth compatible with reduced atomic interactions, we expand the atom cloud over the lattice sites in the fibre by temporarily turning off the lattice confinement for $t_f = 60$ ms, while maintaining the dipole trapping in the radial direction. The time chart outlined in Fig. 1c allows us to extend the cloud length to $l_a = t_f \times 2\sqrt{\langle v_z^2 \rangle} \approx 2.8$ mm, where we use an atomic velocity of $\sqrt{\langle v_z^2 \rangle} \approx 23$ mm s$^{-1}$, estimated from the Doppler width of 55 kHz. This procedure reduces the mean atom occupation from $\bar{m} \approx 1.7$ down to $\bar{m} \approx 0.45$ (see the blue circles in Fig. 3a), while preserving an optical depth of $OD(0) \approx 2.5$. The blue symbols in Figs. 3b and 3c show that the collisional broadening and shift are successfully suppressed by applying this procedure. However, the achieved linewidth of 11 kHz suggests that some unexplained broadening of several kHz still remains.

*Light-polarization-dependent shift:*

To elucidate the source of this residual broadening, we investigate the birefringence of the HC-PCF. In addition, to improve the spatial resolution in the fibre, we reduce the collisional shift by limiting the number of atoms to $N_a < 1200$, i.e., $OD(0) < 1$, instead of expanding atom clouds. Assuming the lattice laser polarization $\epsilon_L$ to be parameterised by an angle $\theta_L$ as defined previously, Eq. (1) becomes $\Delta\nu_L = \Delta\tilde{\alpha}(\lambda_L, \theta_L) I_L$. For the lattice laser wavelength at $\lambda_L = 813$ nm, the differential light shift can be removed by setting $\theta_L = 46°$. The angle sensitivity of the tensor light shift[27] $\frac{d\Delta\tilde{\alpha}}{d\theta_L}\Big|_{\theta_L=46°} = -0.17$ kHz kW$^{-1}$cm$^2$ deg.$^{-1}$ makes the light shift an efficient probe for the fibre birefringence.

Figure 4a shows the lattice-intensity-dependent light shift, where the gradient indicates the effective differential polarizability $\Delta\tilde{\alpha}(\lambda_L, \theta_L) = \Delta\nu_L/I_L$. The blue filled and empty circles are measured for atoms inside ($z_{813} = 3.7$ mm), and outside ($z_0 = -1.6$ mm) the fibre, respectively. While the data confirms that the atomic resonance frequencies are unaffected by being guided in the fibre (as demonstrated by the $I_L \to 0$ extrapolations that coincide at the same frequency), the change of polarizabilities inside ($\Delta\tilde{\alpha}_{in}$) and outside ($\Delta\tilde{\alpha}_{out}$) the fibre indicates the presence of fibre-induced birefringence. Assuming an angle sensitivity of the tensor light shift, $\Delta\tilde{\alpha}_{in} - \Delta\tilde{\alpha}_{out} \approx 0.09$ kHz kW$^{-1}$cm$^2$ corresponds to a polarization rotation of $\delta\theta_L \approx 0.5°$ between $z_0$ and $z_{813}$. We investigate the position-dependent birefringence effect throughout the fibre, which is found to be within $\delta\theta_L \approx 0.3°$ and, in particular, nearly constant for $0 < z < 8$ mm. A relatively large deviation is found in the region of fibre support and clamp (see Fig. 1a), which may indicate the presence of pressure-induced fibre-birefringence. In the following measurements, and in order to be free from stress-induced effects, we focus our attention to the fibre position at around $z \approx 4$ mm.

To moderate fibre-birefringence effect, further experiments are performed at a magic wavelength $\lambda_L = 914$ nm, with $\theta_L = 90°$, where the angle dependence $|d\Delta\tilde{\alpha}/d\theta_L|$ appears only in second order. The filled and empty red circles in Fig. 4a show the resulting reduction in sensitivity measured at $z_{914} = 4.3$ mm and at $z_0 = -1.6$ mm, respectively. The slight change of the position from the measurements at $\lambda_L = 813$ nm results from the lattice wavelength difference, which scales as $\frac{z_{813}-z_0}{z_{914}-z_0} = \frac{813}{914}$, since we use the same detuning sequence $\delta\nu(t)$ for the moving lattice.

In this measurement we simultaneously record the absorption spectra for five different intensity settings so as to minimize the influence of laser frequency drift in extrapolating the lattice intensity $I_L \to 0$. The data points are fit by $\nu_{PCF(FS)} = \Delta\tilde{\alpha}_{PCF(FS)}I_L + y_0$, where $\nu_{PCF(FS)}$ and $\Delta\tilde{\alpha}_{PCF(FS)}$ denote the resonant frequency and the differential polarizability in the PCF or in free space (FS), respectively, and $y_0$ assumes an offset frequency chosen to be zero in Fig. 4a. We evaluate the uncertainty of zero-intensity intercepts by the uncertainty of $y_0$, which are $0.18$ kHz and $0.11$ kHz for 813 nm and 914 nm, respectively, as indicated by error bars at $I_L = 0$. The results indicate that the atomic resonance frequency is unaffected by the fibre with an uncertainty of $\approx 3 \times 10^{-13}$.

Figure 4b shows a spectrum measured at the magic wavelength $\lambda_L = 914$ nm with the lattice intensity of $I_L = 37$ kW cm$^{-2}$ measured at $z \approx 5.3$ mm. The linewidth of $7.8(4)$ kHz agrees well with the saturation-broadened linewidth of $7.8$ kHz for the probe laser intensity $I_\text{p} \approx 0.077 I_0$, demonstrating that there is no significant decoherence of atoms in the fibre at kHz level. At $\lambda_\text{L} = 914$ nm, we investigate the atomic resonance frequencies throughout the fibre, which are found to be within 2 kHz. This variation is partly due to the frequency drift of the probe laser and partly due to the spatial inhomogeneity of the fibre. The detailed investigations of fibre-dependent inhomogeneity, such as, local stress on the fibre, charging effect, and formation of patch potential on inner surface of the fibre are underway.

**Discussion**

The scheme developed here offers a new and ideal platform for high-precision spectroscopy with enhanced signal-to-noise ratio, particularly suitable for the miniaturization of optical lattice clocks operated on the $^1S_0 - {}^3P_0$ clock transitions[30]. The systematic uncertainties of such clocks are essentially characterized by the nuclear spin $I$ of an interrogated isotope, which at the same time decides its quantum statistical nature. Bosonic isotopes[31,37-39], e.g. $^{88}$Sr and $^{174}$Yb, are highly susceptible to collisional shift, therefore, they certainly demand singly-occupied lattices as demonstrated in a 3D optical lattice clock[31]. Recent observations suggest that, as the uncertainty of the clocks approaches $10^{-17}$, collisional interactions become a concern for clocks even with spin-polarized ultracold fermions[40], such as $^{87}$Sr and $^{171}$Yb, where the *s*-wave collisions are suppressed. Our demonstrations of a singly-occupied lattice by expanding atoms in the fibre should be effective for both isotopes to reduce collisional shift while preserving the number of atoms or the QPN limit.

In contrast to free space optics, fibre optics requires special care for the state of light polarization, which is easily affected by mechanical stress or inhomogeneity of the fibre. As the light polarization affects the light shift for the electronic states that have non-zero angular momentum $F \neq 0$, fermionic isotopes with half-integer nuclear spin become susceptible to fibre birefringence even in the clock states with total electronic angular momentum $J = 0$. However, compared to the $^3P_1$ state chosen here as a sensitive probe, the tensor shift in the clock transition of $^{87}$Sr is 7 orders of magnitude smaller[41], as it originates solely from its nuclear spin $I = 9/2$. Our measurements, therefore, suggest that the tensor contribution is safely neglected in achieving $10^{-18}$ clock uncertainty.

A 32-mm-long HC-PCF, as employed here, will support as many as $10^5$ lattice sites or $N_a \approx 10^5$ atoms free from both collisions and light shifts, allowing to achieve a projection-noise-limited stability of $10^{-17}/\sqrt{\tau/s}$, with $\tau$ the averaging time. This is in contrast with free-space lattice-clock experiments that employ $\sim 10^3$ atoms confined in less than 1-mm-long 1D lattices. Further increase of the number of atoms should be possible by extending the fibre length. Moreover, high optical depth and long atomic coherence time allow applying dispersive measurement of atoms[42], quantum non-demolition (QND) measurement protocols, and spin-squeezing of atoms during clock operation[6]. By heterodyning or homodyning the transmittance of a probe laser[8], a quantum feedback scheme[43] may be used to steer the probe laser frequency, instead of applying conventional projection measurements[7]. The strong coupling of atoms to guided modes allows the investigation of collective effects such as collective Lamb shifts[44] and superradiance[45]. In particular, superradiant lasing[46] on the clock transition or generation of narrow-line light source via the phase-matching effect[47] may have potential to replace bulky reference cavities[48,49] required for optical clocks, which will lead to significant miniaturization of optical clocks. Moreover, a fully populated 1D chain of $10^5$ or more qubits sharing an optical bus of the fibre-guided mode could be used for quantum computing and simulation[5] by providing individual spectroscopic access[50] with a magnetic or electric field gradient.

In summary, we have demonstrated precision spectroscopy of atoms in a HC-PCF, investigating possible hurdles intrinsic to fibres, such as collision-limited lifetime, atom-atom interactions, and fibre-induced birefringence. In the present experiment, the coherence time of the system is essentially limited by the natural lifetime of the $^3P_1$ state. Further investigation of the coherence time up to a second is possible by interrogating the $^1S_0 - {}^3P_0$ clock transition[30], which also reduces the sensitivity to fibre birefringence. The novel platform demonstrated here could have an immense impact on future metrology and quantum information sciences using miniaturized atomic devices.

**Methods**

**Frequency shift due to atom fibre-wall interactions**

The Casimir-Polder interaction energy[51] between an atom with polarizability $\alpha$ and an infinite surface at a relatively large distance $r_c$, for which the retardation limit is valid, is given by $U_{CP} = \frac{3\hbar c}{32\pi^2 \epsilon_0 r_c^4} \alpha \Gamma$ with $\epsilon_0$ the vacuum permittivity. The coefficient $\Gamma$ depends on the properties of the surface; $\Gamma = 1$ for ideal metals and $\Gamma < 1$ for dielectric materials. If one considers a photonic crystal with air-filling fraction $p = 0.94$ as a dielectric with relative permittivity $\varepsilon = 1 \cdot p + \varepsilon_{FS} \cdot (1-p)$, then $\Gamma = \frac{\varepsilon-1}{\varepsilon+1} \approx 0.08$ is expected, where we assume the static permittivity of fused silica to be $\varepsilon_{FS} = 3.8$. The energy shift of an atom inside the HC-PCF can be larger than the value given by $U_{CP}$ by a geometric factor $G \sim 6$, which accounts for the atom interaction with six walls (see Fig. 1e) when each of these walls is approximated by an infinite plane.

The difference in polarizabilities for a Sr atom in the $^1S_0$ and $^3P_0$ states[52] is $\Delta\alpha \approx 4 \cdot 10^{-39}$ C m²V⁻¹. For $r_c = 20$ μm, the frequency shift is given by $\Delta\nu \approx \frac{1}{h}\frac{3\hbar c}{32\pi^2\epsilon_0 r_c^4}\Delta\alpha \Gamma G \approx 0.6 \cdot 10^{-3}$ Hz, which corresponds to a fractional clock shift $\Delta\nu/\nu_0 \approx 1.5 \cdot 10^{-18}$. This estimate applies for the zero temperature limit. At room temperature, thermal effects become the same order as the zero fluctuations input. According to Ref. 26, this gives a 3-4 times enhancement, and the atom-wall interaction corresponds to a fractional frequency shift of ~$10^{-17}$.

For the $^1S_0 - {}^3P_1$ transition, the fractional shift can be 20% larger, because of a 20% increase in the differential polarizability $\Delta\alpha$. In this transition, however, the resonant dipole-dipole interaction may be more relevant[53], because of the significantly larger dipole moment than that of the $^1S_0 - {}^3P_0$ clock transition. As the atom-wall distance is much larger than the transition wavelength, $r_c/(\lambda_p/2\pi) \sim 180$, the retardation effect dominates. Considering the current measurement precision of ~$10^{-13}$, which is ~$10^{-2}\gamma_p$ with $\gamma_p = 7.5$ kHz the natural linewidth, the atom-wall interactions can be safely neglected.

**Coupling light into the hollow-core fibre**

The probe and lattice lasers, which are sent through polarization-maintaining single-mode fibres (PM-SMF), are coupled to the HC-PCF using aspheric lens pairs: The output of the PM-SMF is collimated by a $f = 4.6$ mm lens to pass through a vacuum viewport without aberrations and is then matched to the $HE_{11}$ mode of the HC-PCF by a $f = 18.4$ mm lens. Typically 90% of the laser power is transmitted through the 32-mm-long fibre. The far-field intensity pattern is nearly Gaussian as shown in Fig. 1f. The spatial mode after the HC-PCF is verified by recoupling it to another PM-SMF, where we achieve an overall (SMF-HC-SMF) coupling efficiency of 70%.

**HC-PCF design and fabrication**

The fibre is fabricated using the standard stack-and-draw technique. The cladding structure is that of a Kagome-lattice with a pitch of 14 µm (Fig. 5b) and strut thickness of 196 nm (Fig. 5c). This is the smallest silica strut thickness so far reported for a hypocycloid core HC-PCF[33]. This allows the fibre to guide light with low loss for wavelengths as short as 400 nm (see Fig. 5a), and thus covering the experimental operating wavelengths of 813 nm and 689 nm with loss figures of 530 dB km$^{-1}$ and 650 dB km$^{-1}$, respectively.

**HE$_{11}$ mode intensity profile and electric field distribution**

The fibre hollow core has a hypocycloid contour with inner radius $r_c \sim 17$ µm.[33] Figure 6 shows the norm of the two polarization degenerate electric fields of the modes along the two axes of symmetry of the fibre core. The presented simulations are performed over a spectrum of 800-830 nm to cover our operating wavelength of 813 nm, where the mode size has a very moderate change with wavelength. The vertical dashed lines indicate the radial position of the field at $e^{-1}$ of its maximum, corresponding to a mode-field (MF) radius of the HE$_{11}$ mode of ~12.7 µm. This Bessel intensity transverse profile fits to a Gaussian profile with $e^{-2}$ of the maximum radial position at $w_0 = 11.8$ µm. The electric and magnetic fields for the fundamental core-mode HE$_{11}$ are computed using the finite-element-method. Figure 7 shows the components of the electric **E** (V m$^{-1}$) and magnetic **B** (T) fields, when the total optical power contained in the HE$_{11}$ mode is set to 1 W. The results show that the magnitude of the longitudinal component $E_z$ is almost 100 times smaller than the transverse components $(E_x, E_y)$.

**HE$_{11}$ birefringence**

The fabricated fibre core exhibits a small ellipticity, which results in a residual birefringence $\Delta n_{\text{eff}}$. Figure 8 shows the spectrum of the birefringence near the lattice laser wavelength 813 nm. The birefringence is found to be $9.6 \times 10^{-8}$ (i.e. a beat length of 8.4 m), which is more than one order of magnitude lower than the typical photonic bandgap HC-PCF[17]. It is noteworthy that, in addition to the intrinsic fibre form, the birefringence is also induced by mechanical and/or thermal stress. In the case of a photonic bandgap HC-PCF, the lateral pressure-induced birefringence was measured[54] to be in the range of $\partial \Delta n_{\text{eff}}/\partial p \sim 10^{-11}$ Pa$^{-1}$.

**Collision shifts**

We evaluate the collision-shift $\Delta \nu_{\text{col}}(\bar{m})$ as

$$\Delta \nu_{\text{col}}(\bar{m}) = \sum_{k=2}^{\infty} \beta n_1 (k-1) P(k, \bar{m}),$$

where $\beta$ is the collision-shift coefficient, $n_1 = 1/v$ is the atom density for a singly-occupied lattice site with $v = 7.8 \times 10^{-13}$ cm$^3$, and $P(k, \bar{m}) = \frac{\bar{m}^k e^{-\bar{m}}}{k!}$ assumes the Poisson distribution of atoms with mean occupancy $\bar{m}$. The red and blue dashed curves in Fig. 3c show $\Delta \nu_{\text{col}}(\bar{m})$, with $\beta = -1 \times 10^{-9}$ Hz $\cdot$ cm$^3$ to fit the corresponding data points, with red and blue filled circles. This collision-shift coefficient $\beta$ agrees reasonably well with that measured previously by the JILA group[29] $\beta_{\text{JILA}} = -1.3(3) \times 10^{-9}$ Hz $\cdot$ cm$^3$.

**Acknowledgements** This work received support partly from the JSPS through its FIRST Program and from the Photon Frontier Network Program of MEXT, Japan. We thank N. Nemitz for a careful reading of the manuscript. FB acknowledges support from "Agence Nationale de Recherche". FN is partially supported by the RIKEN iTHES Project, MURI Center for Dynamic Magneto-Optics, JSPS-RFBR contract No. 12-02-92100, and Grant-in-Aid for Scientific Research (S).

**Author contributions** HK envisaged and initiated the experiments. HK, TT and SO designed the apparatus and experiments. SO and TT carried out the experiments and analysed the data. SO, TT, and HK discussed experimental results and equally contributed to the experiments. FB and TB designed and fabricated the fibre for the experimental requirements, and LV calculated the fibre modal fields. ZM, VY, and FN calculated the atom-wall interactions, and all authors participated in discussions and the writing of the text.

**Competing financial interests** The authors declare no competing financial interest.

**Correspondence** and requests for material should be addressed to HK (e-mail: katori@amo.t.u-tokyo.ac.jp).

**Figure Legends**

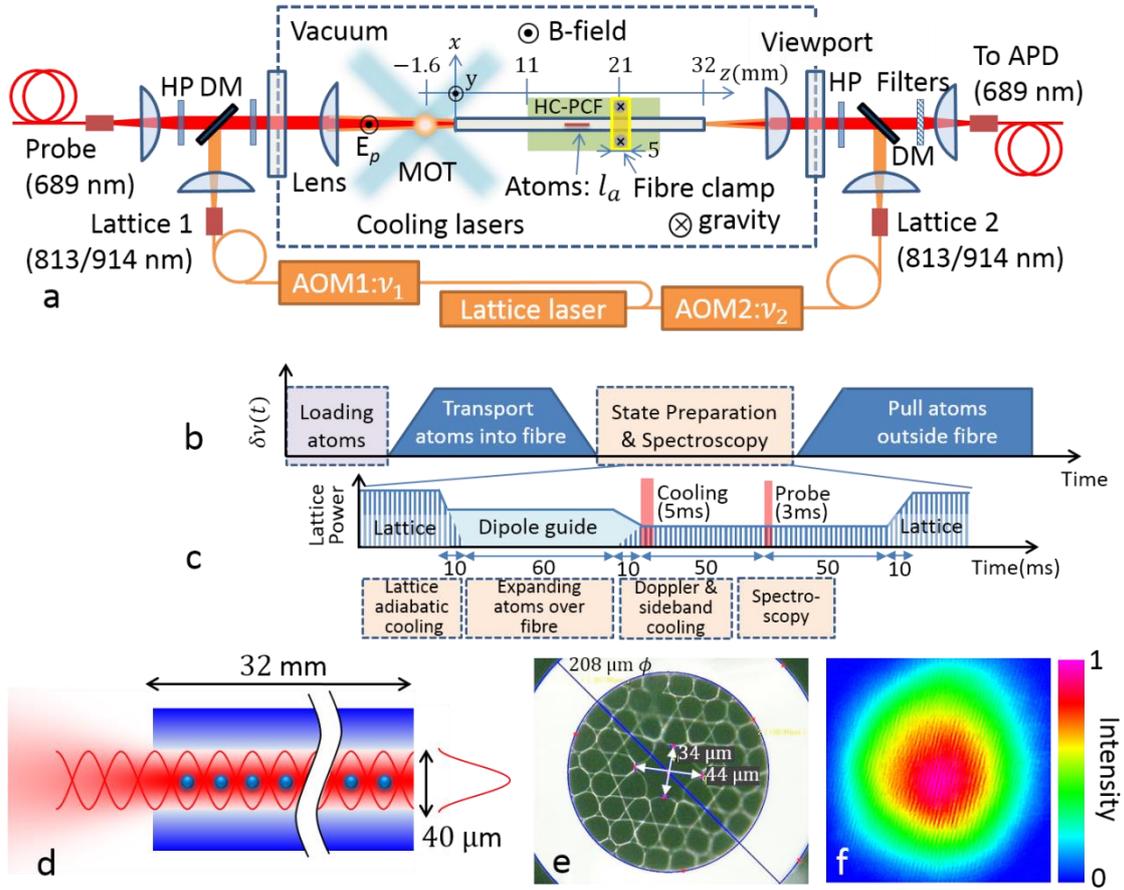

**Figure 1 Experimental setup. a,** Magneto-optically trapped (MOT) atoms are loaded into an optical lattice at around $z = -1.6$ mm from the fibre facet. **b,** By tuning the frequency difference $\delta v(t) = v_2 - v_1$ of the lattice lasers, the atoms are transported into a HC-PCF. A probe laser excites the $^1S_0 - ^3P_1(m = 0)$ transition at 689 nm, with the transmitted light being fed into an avalanche photo-diode (APD) via a single-mode fibre (SMF) after eliminating lattice photons. HP: half-wave plate; DM: dichroic mirror; AOM: acousto-optic modulator. **c,** Atom-expansion protocol. By slowly turning off one of the lattice lasers, we let atoms freely expand along the fibre axis guided by a dipole trap over $l_a = t_f \times 2\sqrt{\langle v_z^2 \rangle} \approx 2.8$ mm within $t_f = 60$ ms. Afterwards, the optical lattice is gradually recovered over 10 ms, while the probe-laser frequency is chirped from $\Delta\nu_p = -35$ kHz to $-70$ kHz in 5 ms, to resettle atoms into lattice sites by Doppler and sideband cooling. **d,** Atoms in the HC-PCF are confined radially and axially by the optical lattice, preventing atoms from interacting with the fibre-wall. **e,** Microscope image of the Kagome fibre. **f,** Far-field pattern of the laser intensity passed through a 32-mm-long HC-PCF.

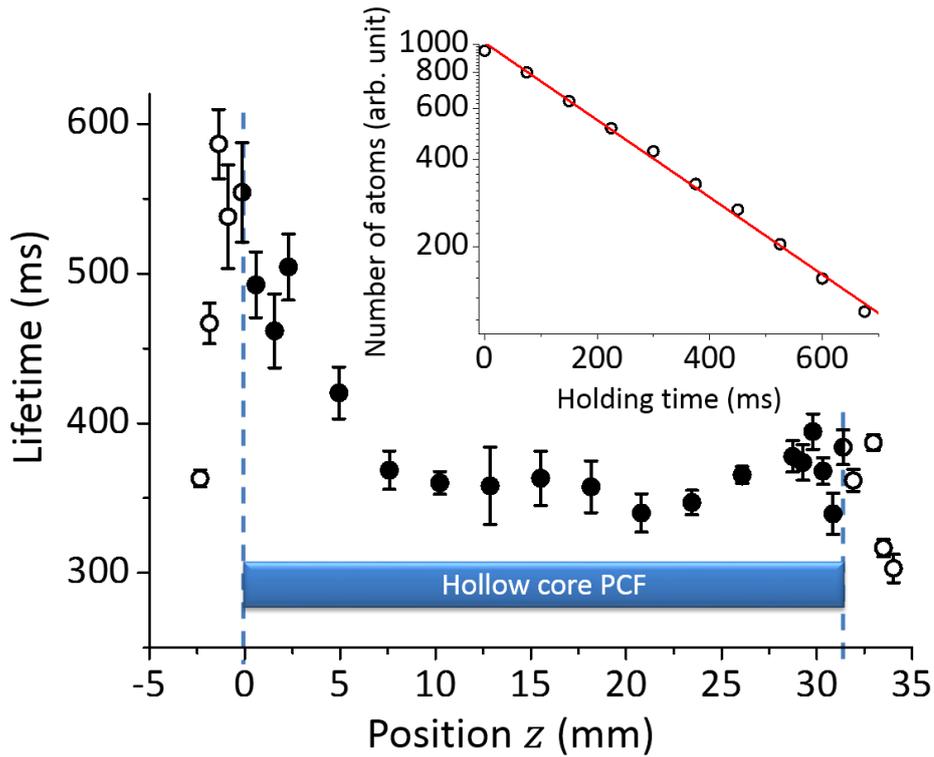

**Figure 2 Lifetime of atoms in a hollow-core fibre.** The lifetime is determined by the number of atoms $N_a$ that pass through the fibre as a function of holding times $\Delta t$ at a given $z$, where $N_a$ is observed by laser-induced fluorescence on the $^1S_0 - {}^1P_1$ transition at the exit of the fibre. The inset shows the decay of the number of atoms (empty circles) measured at $z = 23.4$ mm. The red line shows the exponential fit to data points, which determines the lifetime to be $\tau = 347$ ms $\pm$ 8 ms. After loading atoms from the MOT at $z = -1.6$ mm, the atoms enter the 32-mm-long fibre at $z = 0$ as indicated by the blue region. Empty circles show the lifetimes of atoms outside the fibre, which become longer as the atoms approach the fibre entrance, because of the increase of the potential depth. The lifetimes inside the fibre (filled circles) decrease towards the middle of the fibre due to the increase in collisions with residual gases. Asymmetric behaviour of the position-dependent lifetime with respect to the middle of the fibre ($z = 16$ mm) may be responsible for the excess heating of atoms during transport, which makes lifetimes shorter for a given trap depth and background gas pressure as $z$ increases. Error bars indicate the standard error in the fitting for each data point.

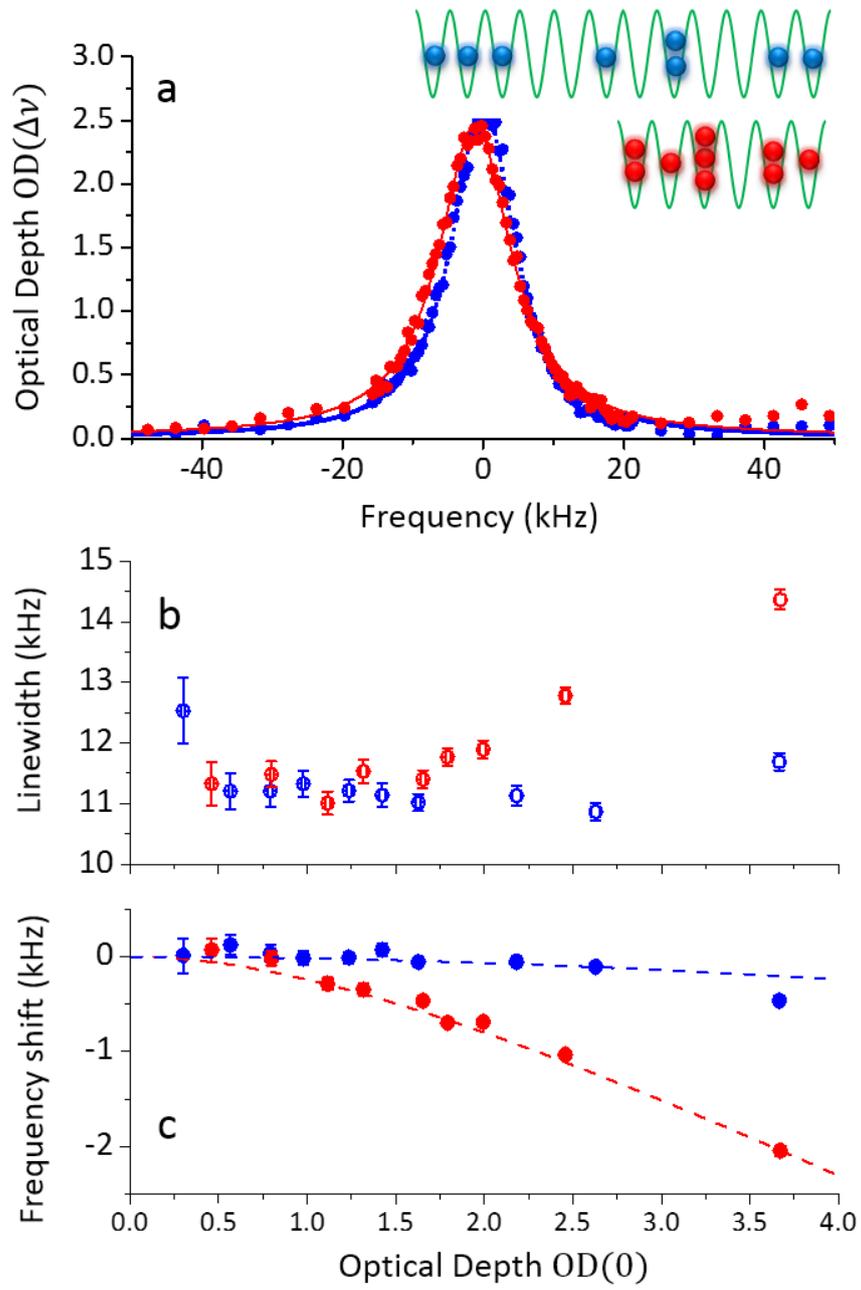

**Figure 3 Spectroscopy of atoms in a fibre. a,** Absorption spectra with and without atomic expansion over lattice sites, as illustrated in the inset, are displayed by blue and red symbols, respectively, corresponding to a mean atom occupation of $\bar{m} \approx 0.45$ and $\bar{m} \approx 1.7$. **b,** Spectral linewidth and **c,** shift as a function of $\mathrm{OD}(0)$. The error bars display the standard error. The atom-number-dependent broadening and shift of the spectrum are suppressed by applying a lattice-expansion protocol, as shown by the blue symbols. Reducing $\mathrm{OD}(0) < 0.8$ for the original atom cloud length $l_\mathrm{a} \approx 700$ μm, which corresponds to $\bar{m} < 0.55$, nearly suppresses collisions (see red circles). The red and blue dashed curves assume a collision shift coefficient of $\beta = -1 \times 10^{-9}$ Hz·cm$^3$. The data points indicated by the dotted rectangle correspond to the spectra shown in **a.**

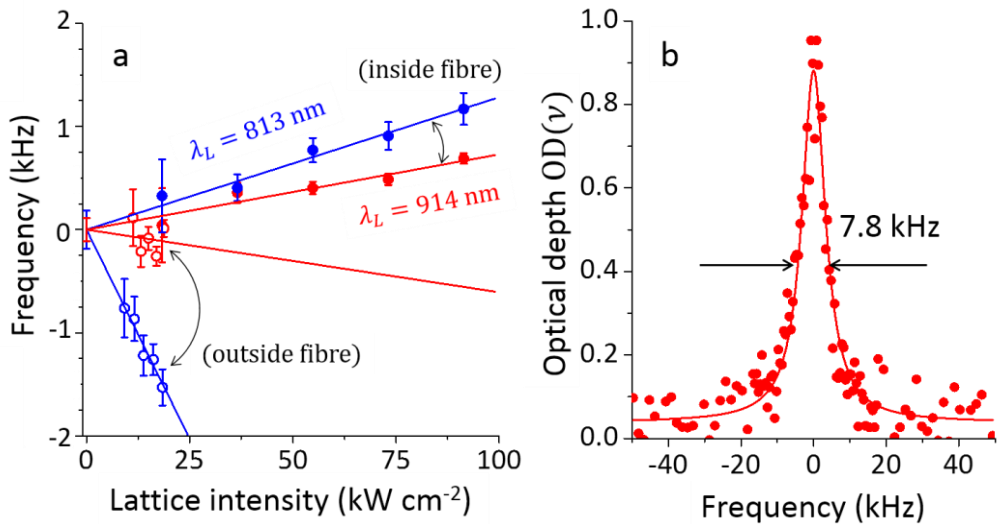

**Figure 4 Measurement of the fibre-induced birefringence effects. a,** Intensity-dependent light shifts on the $^1S_0 - {}^3P_1(m=0)$ transition for atoms inside (filled circles measured at $z = 4.3$ mm and $3.7$ mm) and outside (empty circles measured at $z = -1.6$ mm) the fibre at $\lambda_L = 813$ nm (red) and $914$ nm

(blue) with $\theta_L = 46°$ and $90°$, respectively. For respective wavelengths, the data points are linearly fitted assuming respective $y$-intercepts, $y_0^{813}$ and $y_0^{914}$, which are then taken to zero and their standard errors are indicated by error bars at $I_L = 0$ with respective colours. The fitted gradient is a measure of the differential polarizabilities $\Delta\tilde{\alpha}(\lambda_L, \theta_L) = \Delta\nu_L/I_L$, which are sensitive to the fibre birefringence effects. **b,** A 7.8-kHz-wide spectrum, which agrees with the saturation broadened linewidth for $I_p \approx 0.077 I_0$, is observed for $I_L = 37$ kW cm$^{-2}$ and $\lambda_L = 914$ nm.

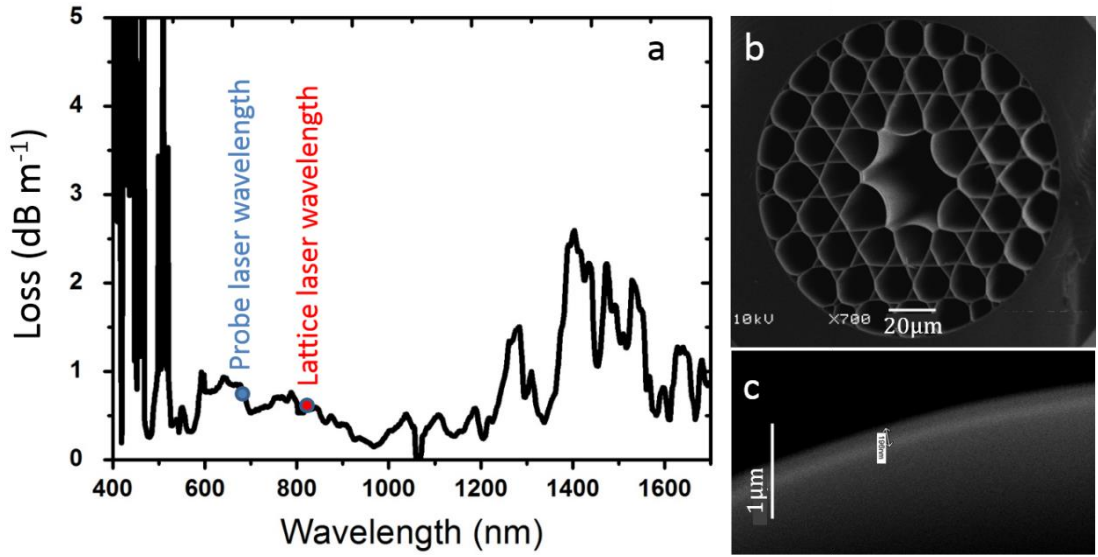

**Figure 5 A Kagome HC-PCF used in the experiment. a,** Loss spectrum of the fibre. **b,** Scanning electron micrograph of the fibre transverse structure. **c,** A zoom-out of the cladding silica strut.

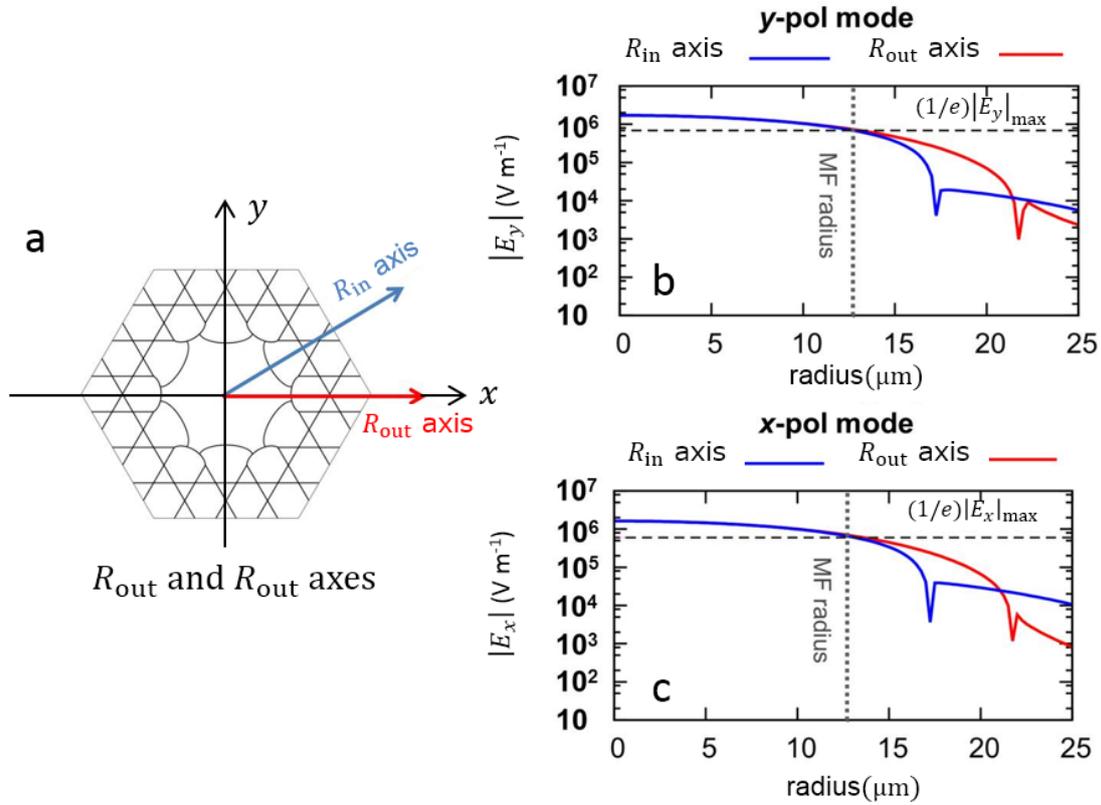

**Figure 6 Transverse profile of the *x*-polarized and *y*-polarized HE$_{11}$ electric field norms along the $R_{\text{in}}$ and $R_{\text{out}}$ axes. a,** Transverse profile of the simulated fibre structure. $R_{\text{in}}$ and $R_{\text{out}}$ are the axes along which the field is calculated. **b and c,** HE$_{11}$ electric field norm profile along the $R_{\text{in}}$ (blue curve) and $R_{\text{out}}$ (red curve) axes for y-polarization (**b**) and x-polarization (**c**). The electric field in $\text{V m}^{-1}$ has been calculated for a total power of 1 W. The mode-field (MF) radius of the HE$_{11}$ mode is ~12.7 µm, as indicated by vertical dashed lines. The horizontal dashed curves indicate the $e^{-1}$ of the maximum field norm.

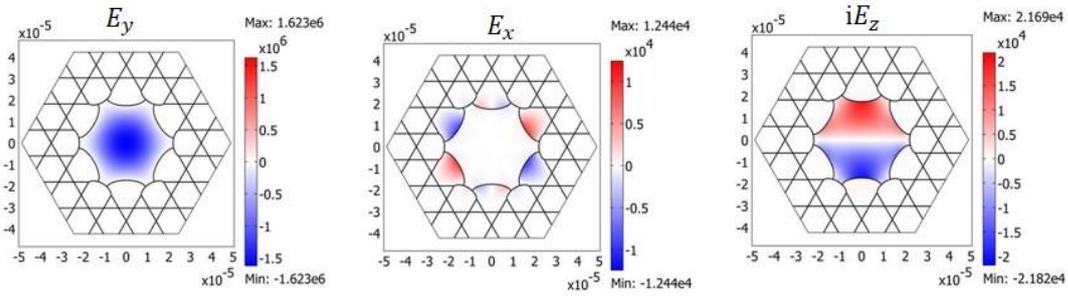
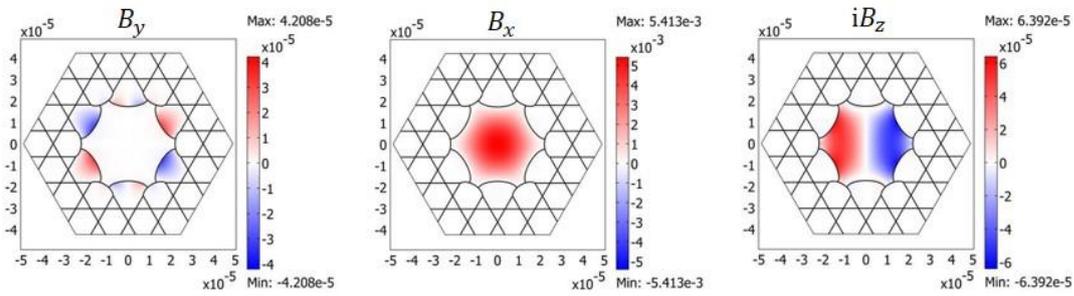

**Figure 7 Electric and magnetic field components of the *y*-polarized HE$_{11}$ mode.** The electric field in $\text{V m}^{-1}$ and magnetic field in T have been calculated for a total power of 1 W.

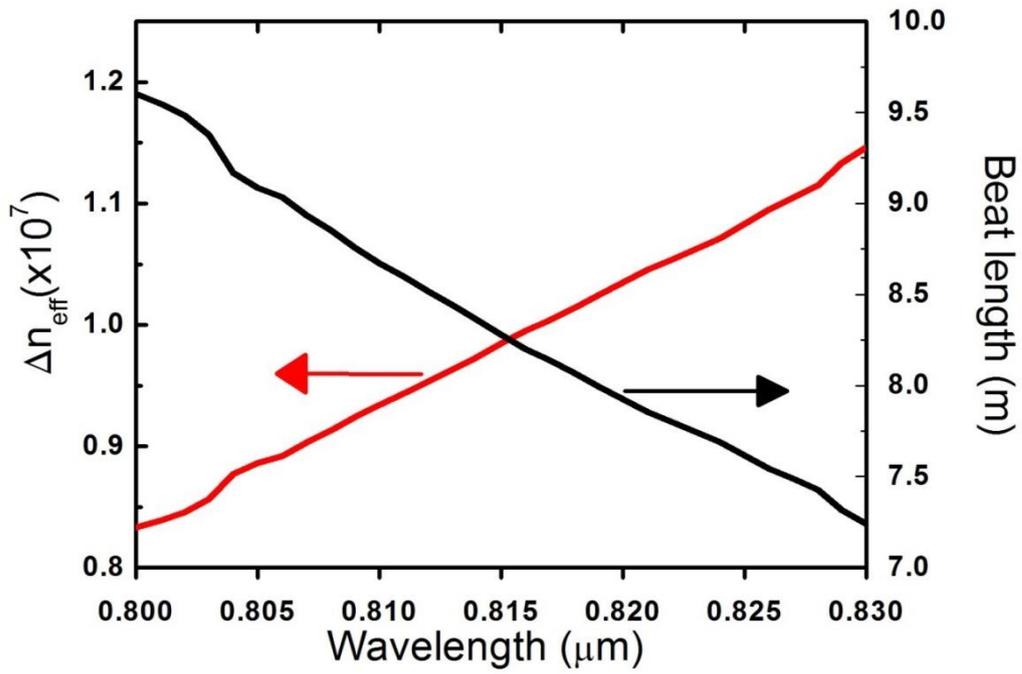

**Figure 8 Spectrum of the $HE_{11}$ birefringence ($\Delta n_{eff}$) and its corresponding beat length.**